\documentclass[11pt]{article}
\usepackage{jheppub}
\usepackage{tikz}
\usetikzlibrary{decorations.pathmorphing}
\tikzset{snake it/.style={decorate, decoration=snake}}
\usetikzlibrary{decorations,decorations.pathmorphing}

\usepackage{graphicx,subcaption}
\newcommand{\dd}{\mathrm{d}}

\title{More on Boundary Holographic Witten Diagrams}

\author[a,b]{Yoshiki Sato}
\affiliation[a]{Department of Physics, Faculty of Science, The University of Tokyo, Bunkyo-ku, Tokyo 113-0033, Japan }
\affiliation[b]{Department of Physics, University of Washington, Seattle, Wa, 98195-1560, USA}

\preprint{\today}

\abstract{In this paper we discuss geodesic Witten diagrams in general holographic conformal field theories with boundary or defect.
In boundary or defect conformal field theory, two-point functions are non-trivial and can be decomposed into conformal blocks in two distinct ways; ambient channel decomposition and boundary channel decomposition.
In our previous work \cite{KS} we only consider two-point functions of same operators.
We generalize our previous work to a situation where operators in two-point functions are different.
We obtain two distinct decomposition for two-point functions of different operators.
}

\begin{document}
\maketitle


\section{Introduction}

In $d$-dimensional conformal field theory (CFT), a conformal symmetry is $SO(d,2)$ and 
two-point functions and three-point functions are completely determined up to prefactors.
Four-point functions are not completely determined by the conformal symmetry and expressed as a function depending on two conformal cross-ratios.
The function depending on conformal cross-ratios can be decomposed by conformal blocks or conformal partial waves which are fundamental objects in CFT.

In the context of the AdS/CFT correspondence, correlation functions are calculated by Witten diagrams \cite{Witten}.
Recently, holographic duals of the conformal blocks, which are called by ``geodesic Witten diagrams," are proposed \cite{Hijano}.
Intermediate points among a bulk-to-bulk propagator and two bulk-to-boundary propagators in Witten diagrams are integrated in a whole region of anti-de Sitter (AdS) spacetime, but similar points in geodesic Witten diagrams are integrated along geodesics connecting two boundary points.
We can confirm that the geodesic Witten diagrams are actually a holographic dual of conformal partial waves by explicit computations.
More elegant way is to check that the geodesic Witten diagrams are eigenfunctions of a Casimir operator with a suitable boundary condition.

The story is changed when boundaries or defects exist in CFT \cite{McAvity}.
The boundaries or defects reduces its conformal symmetry from $SO(d,2)$ to $SO(d-1,2)$ when the boundaries or defects preserve the conformal symmetry.
In this case, one-point functions do not vanish, and two-point functions become non-trivial.
Hence, the two-points functions will be decomposed by conformal blocks instead of four-point functions.
There are two distinct decompositions in boundary CFTs (bCFTs) or defect CFTs (dCFTs);  an ambient channel decomposition and a boundary channel decomposition.
In the ambient channel, its decompositions are based on operator product expansion (OPE)  in a standard ambient spacetime\footnote{As in \cite{ADFK,KS} we use the term ambient spacetime for the $d$-dimensional spacetime whose coordinates are expressed by $x$ labelled by $\mu ,\nu$. The $(d-1)$-dimensional boundary or defect is embedded in the $d$-dimensional spacetime. The direction transverse to the boundary or defect is called $w$ and the directions parallel to them are called $\vec{x}$ labelled by $i,j$. That is, $x = (\vec{x},w)$. 
The term ``bulk" will be reserved for the $(d+1)$-dimensional spacetime of the holographic dual, whose coordinates are $X$ labelled by $M,N$.
We do not use the term ``boundary" as the boundary of AdS spacetime.}.
That is, two-point functions can be expanded by one-point functions of primary operators and their descendants. As mentioned before, one-point functions do not vanish in general because of the presence of the boundary or defect.
The primary operators in the ambient spacetime can be also expanded by local operators localized on the boundary or defect.
Hence, two-point functions are decomposed by two-point functions of boundary operators. This decomposition is called the boundary channel decomposition.

\begin{figure}
\begin{center}
\begin{tikzpicture}
\fill (0,0) circle (3pt);
\draw[line width=1pt] (-3.5,0) --node[above]{$r=+\infty$} (0,0) --node[above]{$r=-\infty$} (3.5,0);
\draw[dashed,line width=1pt] (0,0) -- (0,-3.5);
\draw[dashed,line width=1pt] (0,0) to [out=330,in=175] (340:3.5);
\draw[dashed,line width=1pt] (0,0) to [out=210,in=5] (200:3.5);
\draw[dashed,line width=1pt] (0,0) to [out=300,in=165] (310:3.5);
\draw[dashed,line width=1pt] (0,0) to [out=240,in=15] (230:3.5);
\draw[->] (350:3.8) arc [start angle = 350, end angle = 330, radius = 3.8] ; 
\node (r) at (4,-1.5) [above] {$r$};
\end{tikzpicture}
\caption{A schematic figure of the Janus geometry \eqref{metric}.
The dashed lines are at constant $r$ and represent sliced AdS$_d$ spacetimes.}
\label{fig:metric}
\end{center}
\end{figure}
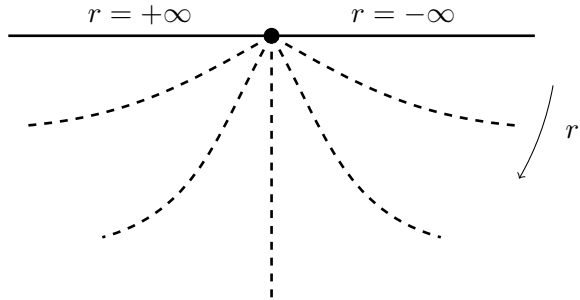

Simple holographic models of bCFTs or dCFTs are described by a $(d+1)$-dimensional metric
\begin{equation}
\dd s^2 = \dd r^2+\mathrm{e}^{2A(r)}\dd s_{\mathrm{AdS}_d}^2 \qquad \text{with} \qquad 
\dd s_{\mathrm{AdS}_d}^2 =L^2\frac{\dd w^2+\dd\vec{x}^2 }{w^2}
\label{metric}
\end{equation}
where $r$ is a radial coordinate and $r\to \pm \infty$ corresponds to an AdS boundary (see Fig. \ref{fig:metric}).
When $\mathrm{e}^{A(r)}=\cosh (r/L)$ (hereafter we set the AdS radius $L=1$), the geometry becomes a pure $(d+1)$-dimensional AdS spacetime.
If one define new coordinates $z$ and $x_d$ as
\begin{equation}
z=\frac{w}{\cosh r} \,, \quad x_d= w \tanh r \,,
\end{equation}
one recover a Poincar\'{e} AdS coordinate.
Examples described by this metric \eqref{metric} include AdS sliced Randall-Sundrum models \cite{KR1,KR2}, the AdS/bCFT proposal by Takayanagi \cite{Takayanagi} and 
Janus solutions of type IIB supergravity \cite{Janus}.
It is possible to generalize \eqref{metric} to a higher dimensional internal space but we do not consider this situation in this paper.

It is naturally expected that the above two decompositions can be realized as geodesic Witten diagrams.
The first attempt has been done by Rastelli and Zhou \cite{RZ}.
They only addressed a simple situation where the defect is a probe brane at $r=0$.
Our previous paper \cite{KS} discussed geodesic Witten diagrams in situations where boundaries or defects are introduced by non-trivial profiles of backgrounds fields on the pure AdS background.
Our situations include the probe brane case \cite{RZ} but addressed the special case where two operators are identical and have the same conformal dimensions for simplicity.
In this paper we generalize our previous work to the case where two operators are different.

The organization of this paper is as follows.
The next section is devoted to preliminaries for the following sections.
We review two distinct decompositions of two-point functions in CFT side and a holographic dual of boundary OPE.
In section \ref{sec3} we decompose of two-point functions with different conformal dimensions into geodesic Witten diagrams.
The ambient channel decomposition and the boundary channel decomposition are 
treated in subsections \ref{subsec31} and \ref{subsec32} respectively.
In section \ref{sec4} we explain how our prescription works in some holographic BCFTs and dCFTs using some examples.
We will conclude in section \ref{con}.

\section{Preliminaries}

\subsection{CFT side}

In this subsection we mainly focus on two-point functions and boundary operator product expansion from the viewpoint of a CFT side.

In CFTs with boundaries or defects, two-point functions are expressed as 
\begin{equation}
\langle \mathcal{O}_1(\vec{x}_1,w_1) \mathcal{O}_2 (\vec{x}_2,w_2) \rangle =
\frac{f(\eta)}{(2w_1)^{\Delta_1}(2w_2)^{\Delta_2}}
\label{eq21}
\end{equation}
where $\eta$ is a conformal cross-ratio,
\begin{equation}
\eta = \frac{(\vec{x}_1-\vec{x}_2)^2+(w_1-w_2)^2}{w_1w_2} \,.
\end{equation}
Note that the two-point functions do not vanish even if the two conformal dimensions $\Delta_1$ and $\Delta_2$ are different.

In the ambient spacetime, the two-point functions can be expanded by a sum of one-point functions using a standard OPE,
\begin{equation}
 \mathcal{O}_1(x_1) \mathcal{O}_2 (x_2) = \sum_k \frac{\lambda_k}{(x_1-x_2)^{\Delta_1+\Delta_2-\Delta_k}}\mathcal{O}_k (x_2) \,.
\end{equation}
This gives us a form of $f$ as
\begin{equation}
f(\eta) = \lambda_{\bf{1}} \left( \frac{4}{\eta} \right)^{\Delta_1}\delta_{\Delta_1,\Delta_2}+
\sum_N \lambda_N a_N f_{\mathrm{ambient}} (\Delta_N, \eta)
\label{24}
\end{equation}
where the sum is over all primary fields and $a$ is determined by an expectation value  of a one-point function,
\begin{equation}
\langle \mathcal{O}(\vec{x},w)\rangle =\frac{a_\mathcal{O}}{(2w)^\Delta} \,.
\end{equation}
We extract a contribution of an identity operator explicitly.
This contribution vanishes when the two conformal dimensions are different.
The contribution of the $N$-th block, $f_{\mathrm{ambient}} (\Delta_N, \eta) $, is given by \cite{Liendo}
\begin{equation}
f_{\mathrm{ambient}} (\Delta_N, \eta)= \left( \frac{\eta}{4} \right) ^{\frac{\Delta_N-\Delta_1-\Delta_2}{2}} {}_2F_1 \left( \frac{\Delta_N+\Delta_1-\Delta_2}{2},\frac{\Delta_N-\Delta_1+\Delta_2}{2},\Delta_N-\frac{d}{2}+1,-\frac{\eta}{4} \right) \,.
\label{26}
\end{equation}

Let us move on to a discussion on the boundary channel.
Primary operators in the ambient spacetime can also be expanded in terms of  boundary localized operators $\mathcal{\hat{O}}(\vec{x})$,
\begin{equation}
\mathcal{O}(\vec{x},w) = \frac{1}{(2w)^\Delta} \sum_k c^\mathcal{O}_{k} (2w)^{\Delta_k} \mathcal{\hat{O}}_{k}(\vec{x})  
\label{bope}
\end{equation}
where 
$\mathcal{\hat{O}}(\vec{x})$ includes both of primaries and their descendants.
This expansion is called boundary operator product expansion (BOPE).
Using BOPE \eqref{bope}, we obtain the boundary channel expansion,
\begin{equation}
f(\eta )=
\sum_{n,m} c_{n}^{\mathcal{O}_1} c_{m}^{\mathcal{O}_2} f_\partial (\Delta_n, \eta)  \delta_{\Delta_n,\Delta_m}
\end{equation}
where the conformal block in the boundary channel is given by \cite{Liendo}
\begin{equation}
f_\partial (\Delta_n, \eta) = \left( \frac{\eta}{4}\right)^{-\Delta_n} {}_2F_1 \left( \Delta_n,\Delta_n-\frac{d}{2} +1, 2\Delta_n-d+2,-\frac{4}{\eta} \right)\,.
\label{29}
\end{equation}

Each conformal blocks \eqref{26} and \eqref{29} satisfy a Casimir equation with a suitable boundary condition.

\subsection{AdS side}

The BOPE can be understood as a mode-decomposition of fields living in the bulk holographically \cite{ADFK}.
A scalar field $\phi_{d+1}(r,x)$ dual to an ambient space operator $\mathcal{O}$ of dimension $\Delta$ satisfies the following equation of motion\footnote{When the scalar field couples to a dilaton field, a term of the dilaton field appears in front of the kinetic term and the equation of motion will be changed.
However, a main story does not so changed.
For example, see our previous paper \cite{KS}.},
\begin{equation}
\label{eom}
(\Box  - M^2(r)) \phi_{d+1} (r,x) = ( {\cal D}^2_r + \mathrm{e}^{-2 A} \partial_d^2 - M^2(r) ) \phi_{d+1} (r,x)= 0 \,. 
\end{equation}
Here $\Box$ is the Laplacian in the full Janus background geometry \eqref{metric} and
$\partial_d^2$ stands for the AdS$_d$ Laplacian on the sliced AdS$_d$ spacetime. The radial operator ${\cal D}^2_r$ is defined as
\begin{equation}
{\cal D}^2_r  := \partial_r^2 + d A'(r) \, \partial_r \,.
\end{equation}
The potential term $M^2(r)$ includes not only a scalar mass $M_0^2$ but also contributions from background fields $X(r)$.
According to the standard AdS/CFT dictionary, the scalar mass is related to the conformal dimension as
\begin{equation}
\Delta (\Delta -d)=M_0^2 \,.
\end{equation}
We make a separation of variables ansatz to the bulk field, 
\begin{equation}
\label{sov}
\phi_{d+1} (r,x) = \sum_n \psi_n(r) \phi_{d,n} (x) \,.
\end{equation}
The separated fields on sliced AdS$_d$ spacetime satisfy 
\begin{equation}
\label{onslicefree}
\partial_d^2 \phi_{d,n} = m_n^2 \phi_{d,n} \,. 
\end{equation}
The eigenvalues $m_n^2$ are then determined by the internal mode equation:
\begin{equation}
\label{215}
{\cal D}^2_r \psi _n(r)  - M^2(r) \psi_n(r) = - \, \, \mathrm{e}^{-2 A(r)} m_n^2 \psi_n(r) \, .
\end{equation}
This second differential equation can be converted to a standard Schr\"odinger equation by a change of variables and a redefinition of the field.
A change of the coordinate from $r$ to a conformal coordinate $z$ with $\mathrm{d}r=\mathrm{e}^A \mathrm{d}z$ removes the $\mathrm{e}^{-2A}$ factor in front of the eigenvalues $m_n^2$ and a redefinition of the field $\psi_n = \mathrm{e}^{-(d-1)A/2} \Psi_n$ removes the first derivatives acting on the modefunction.
Finally the differential equation reduces to a standard Schr\"odinger equation for $\Psi_n(z)$ together with an effective potential \cite{DFGK}
\begin{equation}
\label{potential}
V(z) = \frac{1}{2} \left [ \left ( \frac{d-1}{2} \frac{\mathrm{d}A}{\mathrm{d}z} \right )^2 + \frac{d-1}{2} \frac{\mathrm{d}^2A}{\mathrm{d}z^2} + M^2 \mathrm{e}^{2A} \right ]\, .
\end{equation}
A completeness relation and an orthogonality relation for the original field $\psi_n(r)$,
\begin{equation}
\sum_n \psi_n(r) \psi_n(r') = \mathrm{e}^{-(d-2)A(r)} \delta(r-r') \, , \qquad  \int \! \mathrm{d}r \, \mathrm{e}^{(d-2) A(r)} \, \psi_m \psi_n = \delta_{mn}\,, 
\end{equation}
become standard ones for the rescaled field $\Psi _n(z)$.

According to \cite{ADFK}, the primaries appearing in the BOPE \eqref{bope} are related to the modes $\phi_{d,n}$ in one-to-one correspondence and their conformal dimensions are related with eigenvalues $m_n^2$ as 
\begin{equation}
\Delta_n [ \Delta_n-(d-1) ] = m_n^2 \,.
\end{equation}
A boundary condition of the modefunctions at $r \to \pm \infty$ requires 
\begin{equation}
\Delta_n=\Delta +n \,.
\end{equation}

Finally we summarize a bulk-to-bulk propagator and a bulk-to-boundary propagator to fix our notation and also give a useful decomposition of the bulk-to-bulk propagator.
The bulk-to-bulk propagator is a Green function of the differential operator $\Box -M^2(r)$,
\begin{equation}
(\Box -M^2(r_1)) G_{\Delta,d+1} (r_1,\vec{x}_1,w_1,r_2,\vec{x}_2,w_2)= \frac{1}{\sqrt{-g}}
\delta (r_1-r_2) \delta (\vec{x}_1-\vec{x}_2) \delta (w_1-w_2) \,.
\label{221}
\end{equation}
The bulk-to-boundary propagator is defined as a solution of the following differential equation,
\begin{equation}
(\Box -M^2(r)) K_{\Delta,d+1} (r,\vec{x}_1,w_1,\vec{x}_2,w_2) =0
\end{equation}
and is required to approach an appropriate delta function,
\begin{equation}
\lim_{r\to \infty } \mathrm{e}^{(d-\Delta)r} K_{\Delta,d+1} (r,\vec{x}_1,w_1,\vec{x}_2,w_2)= \delta (\vec{x}_1-\vec{x}_2) \delta (w_1-w_2) \,.
\end{equation}
The bulk-to-boundary propagator is related with the bulk-to-bulk propagator,
\begin{equation}
K_{\Delta,d+1}(r_1,\vec{x}_1,w_1,\vec{x}_2,w_2)= \lim_{r_2 \to \infty} 
(2\Delta -d) \frac{\mathrm{e}^{\Delta r_2}}{(2w_2)^{\Delta}}  G_{\Delta,d+1} (r_1,\vec{x}_1,w_1,r_2,\vec{x}_2,w_2) \,.
\label{bulkboundary}
\end{equation}

The bulk-to-bulk propagator in the Janus metric \eqref{metric} can be decomposed into a sum of products of modefunctions and bulk-to-bulk propagators on the sliced AdS$_d$ spacetime, 
\begin{equation}
G_{\Delta,d+1}(X_1,X_2)=\sum_n \psi_n(r_1) \psi_n(r_2) G_{\Delta_{n},d} (x_1,x_2)  
\label{224}
\end{equation}
with 
\begin{equation}
\Delta_{n}=\Delta+n\,,  \quad n \in \mathbb{N}\,.
\end{equation}
$G_{\Delta_{n},d}$ is a bulk-to-bulk propagator on sliced AdS$_d$ spacetime,
\begin{equation}
G_{\Delta_{n},d} (x_1,x_2) =\frac{C_{\Delta_n ,d}}{2^{\Delta_n}(2\Delta_n-(d-1))}\xi^{\Delta_n} {}_2F_1
\left( \frac{\Delta_n}{2},\frac{\Delta_n+1}{2},\Delta_n-\frac{d-1}{2}+1,\xi^2  \right)
\end{equation}
with
\begin{equation}
C_{\Delta_n ,d}=\frac{\Gamma (\Delta_n)}{\pi^{(d-1)/2}\Gamma (\Delta_n-(d-1)/2)}
\end{equation}
and $\xi$ the chordal coordinate,
\begin{equation}
\xi = \frac{2w_1 w_2}{w_1^2+w_2^2+(\vec{x}_1-\vec{x}_2)^2} \,.
\end{equation}
Using \eqref{bulkboundary}, similar decompositions can be applied to bulk-to-boundary propagators. 
Note that the bulk-to-bulk propagator on sliced AdS$_d$ spacetime is equal to conformal blocks in boundary channel up to prefactor.

\section{Decompositions of two-point function}
\label{sec3}

In this section we decompose two-point functions into geodesic Witten diagrams in a situation where a boundary or a defect is introduced weakly so that the decompositions into geodesic Witten diagrams make sense. 
This situation is achieved by small background fields $X(r)$,
\begin{equation}
X(r) = \varepsilon \, \delta X(r)
\end{equation}
where $\varepsilon$ a small parameter.
Since an energy-momentum tensor constructed from the background fields is corrected at 
second order of $\varepsilon$, the Janus metric \eqref{metric} can be expanded as
\begin{equation}
g =g_{0}+\varepsilon ^2 \, \delta g
\end{equation}
where $g_0$ represents a metric of the pure AdS$_{d+1}$ geometry.
Hence, we can treat the geometry as the pure AdS at leading order of $\varepsilon$.
In the following sections, bulk-to-bulk and bulk-to-boundary propagators are those of the pure AdS. 

First, let us consider a situation where the boundary or defect is absent.
In this case, the two-point function is determined by the conformal symmetry
\begin{equation}
\langle \mathcal{O}_1 (\vec{x}_1,w_1) \mathcal{O}_2 (\vec{x}_2,w_2) \rangle =
\frac{\mathcal{N} \delta_{\Delta_1,\Delta_2} }{[(\vec{x}_1-\vec{x}_2)^2+(w_1-w_2)^2]^{\Delta_1}}
\end{equation}
up to a prefactor $\mathcal{N}$.
Comparing with \eqref{eq21}, we obtain
\begin{equation}
f (\eta)=\mathcal{N} \left( \frac{\eta}{4}\right)^{-\Delta_1} \delta_{\Delta_1,\Delta_2} \,.
\label{35}
\end{equation}
Equation \eqref{35} corresponds to the ambient channel expansion because one-point functions vanish except the identity operator and \eqref{24} reduces to \eqref{35}.
The two distinct decompositions should be same, then the following relation must hold,
\begin{equation}
\mathcal{N} \left( \frac{\eta}{4}\right)^{-\Delta} = \sum_{n=0}^\infty (c_n^{\mathcal{O}})^2 f_\partial (\Delta +n, \eta) \,.
\end{equation}
The BPOE coefficients are obtained as \cite{KS}
\begin{equation}
(c_n^\mathcal{O})^2=(C_n)^2 (2\Delta -d)^2 \frac{C_{\Delta_n,d}}{2\Delta_n-(d-1)}\frac{1}{4^{\Delta_n}}
\end{equation}
where $C_n$ is an  asymptotic value of the modefunction,
\begin{equation}
\psi_n (r) = C_n \mathrm{e}^{-\Delta r} +\mathcal{O} (\mathrm{e}^{-(\Delta+2)r}) \,.
\end{equation}
This identity is a special case of (A.7b) in \cite{Hogervorst} because (A.7b) in \cite{Hogervorst} with $h=d/2,\ell_1=\ell_2 =-\Delta$ and $\rho =\eta /(\eta+4)$ reduces to this identity. Similar identities have also been used in \cite{HH}.
Summarizing the above discussion, two decompositions are the same when the boundary or defect is absent.

Next, we add the boundary or defect as a perturbation and decompose two-point functions in two different ways.
Our claim is that a leading correction to the two-point functions with different conformal dimensions is given by
\begin{equation}
\begin{split}
&\delta \langle \mathcal{O}_{\Delta_1} (\vec{x}_1,w_1) \mathcal{O}_{\Delta_2} (\vec{x}_2,w_2) \rangle \\
& = 2 \varepsilon \int \! \dd ^{d+1}Y' \, \sqrt{-g_0}\,  \mathrm{e}^{-2A} K_{\Delta_1,d+1} (\vec{x}_1,w_1,Y') K_{\Delta_2,d+1} (\vec{x}_2,w_2,Y') \, \delta V_{\Delta_1,\Delta_2} (r')\,.
\label{2pt}
\end{split}
\end{equation}
Here $K_{\Delta_1,d+1}$ and $K_{\Delta_2,d+1}$ are bulk-to-boundary propagators 
between boundary points inserted the operators and a bulk point where a source term 
 $2 \varepsilon \, \mathrm{e}^{-2A} \delta V_{\Delta_1,\Delta_2}$ is inserted. 
The potential is obtained from a variation of the action\footnote{The awkward factor $2 \mathrm{e}^{-2A}$ is included so that this potential reduces a leading correction of \eqref{potential} for $\Delta_1=\Delta_2$. Note that the change of variables and rescaling of the field are required.}
\begin{equation}
2 \varepsilon \, \mathrm{e}^{-2A} \delta V_{\Delta_1,\Delta_2} = -\frac{\delta^2 S}{\delta \phi_{\Delta_1} \delta \phi_{\Delta_2} } \,.
\end{equation}
For $\Delta_1=\Delta_2$, \eqref{2pt} can be obtained explicitly by expanding the differential equation for the bulk-to-bulk propagator \eqref{221} around backgrounds, multiplying it by  another bulk-to-bulk propagator and integrating the intersect \cite{KS}.
We expect that the similar equation of \eqref{2pt} with the same conformal dimensions can be generalized to a case with different conformal dimensions.

In the following two subsections, we give two different decompositions of the two-point function starting from \eqref{2pt}.

\subsection{Ambient channel}
\label{subsec31}

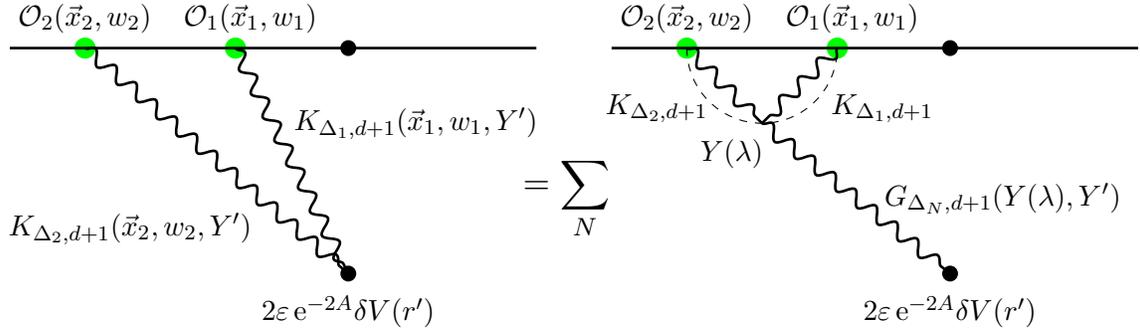
\begin{figure}
\begin{center}
\begin{tikzpicture}
\fill (0,0) circle (3pt);
\fill[green] (-1.5,0) circle (4pt);
\fill[green] (-3.5,0) circle (4pt);
\node[yshift=2pt,above] (O1) at (-1.3,0) {$\mathcal{O}_1(\vec{x}_1,w_1)$};
\node[yshift=2pt,above] (O1) at (-3.5,0) {$\mathcal{O}_2(\vec{x}_2,w_2)$};
\draw[line width=1pt] (-4.5,0) -- (0,0) -- (2.5,0);

\draw[snake it,line width=1pt] (-1.5,0) -- (0,-3);
\draw[snake it,line width=1pt] (-3.5,0) -- (0,-3);
\node at (0.9,-1) {$K_{\Delta_1,d+1}(\vec{x}_1,w_1,Y')$};
\node at (-2.9,-2.4) {$K_{\Delta_2,d+1}(\vec{x}_2,w_2,Y')$};
\node at (0,-3.5) {$2\varepsilon \, \mathrm{e}^{-2A} \delta V(r')$};
\fill (0,-3) circle (3pt);

\node at (2.9,-2) {\scalebox{1.2}{$=\displaystyle \sum_N$}};

\fill (8,0) circle (3pt);
\fill[green] (6.5,0) circle (4pt);
\fill[green] (4.5,0) circle (4pt);
\node[yshift=2pt,above] (O1) at (6.7,0) {$\mathcal{O}_1(\vec{x}_1,w_1)$};
\node[yshift=2pt,above] (O1) at (4.5,0) {$\mathcal{O}_2(\vec{x}_2,w_2)$};
\draw[line width=1pt] (3.5,0) -- (8,0) -- (10.5,0);

\draw[dashed] (4.5,0) arc [start angle = 180, end angle = 360, radius = 1];

\draw[snake it,line width=1pt] (6.5,0) -- (5.5,-1);
\draw[snake it,line width=1pt] (4.5,0) -- (5.5,-1);
\draw[snake it,line width=1pt] (5.5,-1) -- (8,-3);
\node at (7.1,-0.8) {$K_{\Delta_1,d+1}$};
\node at (4.1,-0.8) {$K_{\Delta_2,d+1}$};
\node at (5.1,-1.4) {$Y(\lambda)$};
\node at (8.7,-2) {$G_{\Delta_N,d+1}(Y(\lambda),Y')$};
\node at (8,-3.5) {$2\varepsilon \, \mathrm{e}^{-2A} \delta V(r')$};
\fill (8,-3) circle (3pt);
\end{tikzpicture}
\caption{Schematic picture of the ambient channel decomposition. The dashed line is a geodesic. The point $r=r'$ where the source is inserted is integrated.}
\label{fig:ambient}
\end{center}
\end{figure}

In the ambient channel, we can use similar techniques used in geodesic Witten diagrams without boundary or defect.
The two bulk-to-boundary propagators intersected at a same bulk point can be decomposed as \cite{Hijano}
\begin{equation}
\begin{split}
&K_{\Delta_1,d+1}(\vec{x}_1,w_1,Y')  K_{\Delta_2,d+1}(Y',\vec{x}_2,w_2)  \\
&=\sum_N b_N \int _{\gamma} \! \mathrm{d} \lambda  \,
K_{\Delta_1,d+1}(\vec{x}_1,w_1,Y(\lambda))  K_{\Delta_2,d+1}(Y(\lambda ),\vec{x}_2,w_2)
G_{\Delta_N,d+1}(Y(\lambda),Y')
\label{39}
\end{split}
\end{equation}
with coefficients
\begin{equation}
b_N=
\frac{2\Gamma (\Delta_N)}{\Gamma ((\Delta_N +\Delta_1-\Delta_2 )/2)\Gamma ((\Delta_N -\Delta_1+\Delta_2 )/2)}
\frac{(-1)^N}{N!} \frac{(\Delta_1)_N(\Delta_2)_N}{(\Delta_1+\Delta_2 +N-d/2)_N}  
\end{equation}
where $\Delta_N=2\Delta +2N$ and $(x)_n=\Gamma (x+n)/\Gamma (x)$ is the Pochhammer symbol.
$\gamma$ parameterized by $\lambda$ represents a geodesic anchored to two boundary points $(\vec{x}_1,w_1)$ and $(\vec{x}_2,w_2)$.
Inserting this decomposition \eqref{39} into \eqref{2pt}, the two-point function is decomposed as a sum of geodesic Witten diagrams,
\begin{equation}
\begin{split}
& \delta \langle \mathcal{O}_{\Delta_1} (\vec{x}_1,w_1) \mathcal{O}_{\Delta_2} (\vec{x}_2,w_2) \rangle = 
2 \varepsilon \sum_N b_N \int \! \dd ^{d+1}Y' \, \sqrt{-g_0}\, \mathrm{e}^{-2A}  \\
& \times
 \int _{\gamma} \! \mathrm{d} \lambda  \,
K_{\Delta_1,d+1}(\vec{x}_1,w_1,Y(\lambda))  K_{\Delta_2,d+1}(Y(\lambda ),\vec{x}_2,w_2)
G_{\Delta_N,d+1}(Y(\lambda),Y') \, \delta V_{\Delta_1,\Delta_2} (r') \,.
\label{ambient}
\end{split}
\end{equation}
This is just the ambient channel decomposition (See Fig. \ref{fig:ambient}).

It is difficult to prove these geodesic Witten diagrams are conformal blocks by explicit computations.
However, we can easily confirm this using Casimir equation.
In fact a combination 
\begin{equation}
\int _{\gamma} \! \mathrm{d} \lambda  \,
K_{\Delta_1,d+1}(\vec{x}_1,w_1,Y(\lambda))  K_{\Delta_2,d+1}(Y(\lambda ),\vec{x}_2,w_2)
G_{\Delta_N,d+1}(Y(\lambda),Y')
\end{equation}
satisfy Casimir equation with a suitable boundary condition \cite{RZ}.

\subsection{Boundary channel}
\label{subsec32}

For a purpose of a decomposition in the boundary channel, the decomposition of the bulk-to-bulk propagator on AdS$_{d+1}$ \eqref{224} is useful,
\begin{align}
G_{\Delta_1,d+1}(X_1,Y')&=\sum_n \psi_n^{(1)}(r_1) \psi_n^{(1)}(r') G_{\Delta_{1,n},d} (x_1,y') \,, \\
G_{\Delta_2,d+1}(X_2,Y')&=\sum_m \psi_m^{(2)}(r_2) \psi_m^{(2)}(r') G_{\Delta_{2,m},d} (x_2,y') 
\end{align}
with 
\begin{equation}
\Delta_{1,n}=\Delta_1+n\,, \quad \Delta_{2,m}=\Delta_2+m \quad \mbox{with} \quad n,m \in \mathbb{N}\,.
\end{equation}
To remind us that each mode functions depend on the conformal dimension, the superscripts $(1)$ and $(2)$ are added.
Plugging the corresponding decompositions of the bulk-to-boundary propagators obtained from the relation \eqref{bulkboundary} into \eqref{2pt}, the decomposition of the two-point function includes a sum, 
\begin{equation}
\sum_{n,m} C_n^{(1)}C_m^{(2)} \delta V_{\Delta_1,\Delta_2}(r') \psi_n^{(1)}(r') \psi_m^{(2)}(r') G_{\Delta_{1,n},d} (x_1,y') G_{\Delta_{2,m},d} (x_2,y')
\end{equation}
where $C_n^{(1)}$ and $C_m^{(2)}$ are asymptotic values of the modefunctions.
We decompose this sum whether $\Delta_{1,n}=\Delta_{2,m}$ or not,
\begin{align}
&\sum_{n,m} C_n^{(1)}C_m^{(2)} \delta V_{\Delta_1,\Delta_2}(r') \psi_n^{(1)}(r') \psi_m^{(2)}(r') G_{\Delta_{1,n},d} (x_1,y') G_{\Delta_{2,m},d} (x_2,y') \notag \\
&=\! \sum_{\substack{n,m\\ \Delta_{1,n} \neq \Delta_{2,m} }} \! \! \! \!  \! C_n^{(1)}C_m^{(2)}
\frac{\delta V_{\Delta_1,\Delta_2}(r') \psi_n^{(1)}(r') \psi_m^{(2)}(r') }{(m_n^{(1)})^2-(m_m^{(2)})^2} \left( (m_n^{(1)})^2-(m_m^{(2)})^2 \right) G_{\Delta_{1,n},d} (x_1,y') G_{\Delta_{2,m},d} (x_2,y') \notag \\
&+\!  \sum_{\substack{n\\ \Delta_{1,n} = \Delta_{2,m} }} \! \! \! \!  \! C_n^{(1)}C_m^{(2)} \delta V_{\Delta_1,\Delta_2}(r') \psi_n^{(1)}(r') \psi_m^{(2)}(r') G_{\Delta_{1,n},d} (x_1,y') G_{\Delta_{2,m},d} (x_2,y') \,.
\end{align}
The first sums over $n$ and $m$ such that $\Delta_{1,n} \neq \Delta_{2,m}$ while the second term sums over $n$ such that $\Delta_{1,n} = \Delta_{2,m}$ (Since $m$ is determined by $n$ from the condition $\Delta_{1,n} = \Delta_{2,m}$, a sum over $m$ does not appear).
Hence, the second term appears only in the case where $\Delta_1-\Delta_2$ is integer.
By integrating the first term about $y'$, it becomes
\begin{equation}
\begin{split}
&\int \! \dd ^d y' \sqrt{-g_{\mathrm{AdS}_d}} \left( (m_n^{(1)})^2-(m_m^{(2)})^2 \right) G_{\Delta_{1,n},d} (x_1,y') G_{\Delta_{2,m},d} (x_2,y') \\
&=  G_{\Delta_{1,n},d} (x_1,x_2) - G_{\Delta_{2,m},d} (x_1,x_2) 
\end{split}
\end{equation}
where we use a relation
\begin{equation}
(m_n^{(1)})^2 G_{\Delta_{1,n},d} (x_1,y') = \partial_d^2 G_{\Delta_{1,n},d} (x_1,y') -\frac{1}{\sqrt{-g_{\mathrm{AdS}_d}}} \delta (x_1-y')
\end{equation}
and a similar relation for $G_{\Delta_{2,m},d}$.

\begin{figure}[t]
\begin{center}
\begin{tikzpicture}
\fill (0,0) circle (3pt);
\fill[green] (-1.5,0) circle (4pt);
\fill[green] (-3.5,0) circle (4pt);
\node[yshift=2pt,above] (O1) at (-1.3,0) {$\mathcal{O}_1(\vec{x}_1,w_1)$};
\node[yshift=2pt,above] (O1) at (-3.5,0) {$\mathcal{O}_2(\vec{x}_2,w_2)$};
\draw[line width=1pt] (-4.5,0) -- (0,0) -- (2.5,0);

\draw[snake it,line width=1pt] (-1.5,0) -- (0,-3);
\draw[snake it,line width=1pt] (-3.5,0) -- (0,-3);
\node at (0.9,-1) {$K_{\Delta_1,d+1}(\vec{x}_1,w_1,Y')$};
\node at (-2.9,-2.4) {$K_{\Delta_2,d+1}(\vec{x}_2,w_2,Y')$};
\node at (0,-3.5) {$2\varepsilon \, \mathrm{e}^{-2A} \delta V(r')$};
\fill (0,-3) circle (3pt);

\node at (2.9,-2) {\scalebox{1.2}{$=\displaystyle \sum_n$}};

\fill (8,0) circle (3pt);
\fill[green] (6.5,0) circle (4pt);
\fill[green] (4.5,0) circle (4pt);
\node[yshift=2pt,above] (O1) at (6.7,0) {$\mathcal{O}_1(\vec{x}_1,w_1)$};
\node[yshift=2pt,above] (O1) at (4.5,0) {$\mathcal{O}_2(\vec{x}_2,w_2)$};
\draw[line width=1pt] (3.5,0) -- (8,0) -- (10.5,0);

\draw[dashed] (4.5,0) arc [start angle = 180, end angle = 310, radius = 3.5];
\draw[dashed] (6.5,0) arc [start angle = 180, end angle = 360, radius = 1.5];

\draw[snake it,line width=1pt] (6.5,0) -- (7.25,-1.3);
\draw[snake it,line width=1pt] (4.5,0) -- (6.25,-3.03);
\draw[snake it,line width=1pt] (6.25,-3.03) --  (7.25,-1.3);
\fill (7.25,-1.3) circle (2pt);
\fill (6.25,-3.03) circle (2pt);
\node at (8,-0.7) {$K_{\Delta_1,d+1}$};
\node at (4.6,-1.8) {$K_{\Delta_2,d+1}$};
\node at (7.8,-2.4) {$G_{\Delta_n,d}$};
\node at (8.5,-1.3) {$(\vec{x}_1,w_1,r')$};
\node at (5.8,-3.5) {$(\vec{x}_2,w_2,r')$};
\end{tikzpicture}
\caption{Schematic picture of the boundary channel decomposition. The dashed lines are geodesics. We need to integrate about the point $r=r'$ where the source is inserted.
Anomalous dimension terms are omitted.}
\label{fig:boundary}
\end{center}
\end{figure}
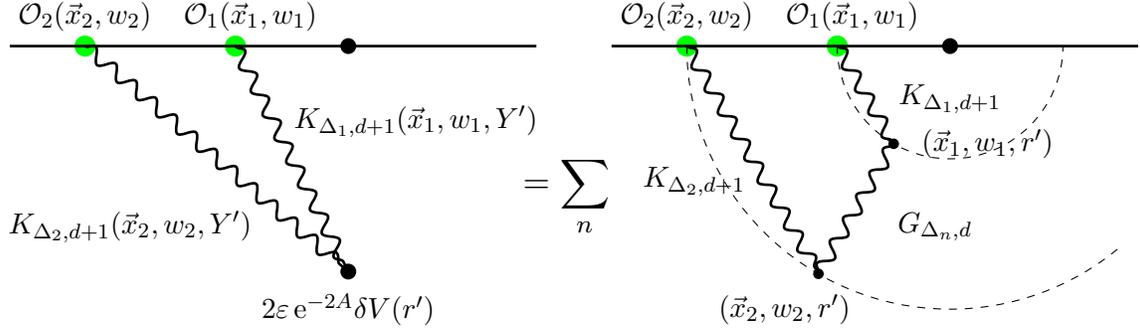

Summarizing the above equations, the two-point function can be decomposed as 
\begin{align}
&\delta \langle \mathcal{O}_1 (\vec{x}_1,w_1) \mathcal{O}_2 (\vec{x}_2,w_2)  \rangle
=2 \varepsilon \frac{(2\Delta_1 -d)(2\Delta_2-d)}{(2w_1)^{\Delta_1}(2w_2)^{\Delta_2}}  \notag \\
&\times \left[  \sum_{\substack{n,m\\ \Delta_{1,n} \neq \Delta_{2,m} }} \! \! \! \!  \! C_n^{(1)}C_m^{(2)} \int \! \dd r' \, \mathrm{e}^{(d-2)A}
\frac{\delta V_{\Delta_1,\Delta_2}(r') \psi_n^{(1)}(r') \psi_m^{(2)}(r') }{(m_n^{(1)})^2-(m_m^{(2)})^2}  \left( G_{\Delta_{1,n},d} (x_1,x_2)- G_{\Delta_{2,m},d} (x_1,x_2)\right) \right. \notag \\
&\left. \qquad \qquad  +\!  \sum_{\substack{n\\ \Delta_{1,n} = \Delta_{2,m} }} \! \! \! \!  \! C_n^{(1)}C_m^{(2)} \int \! \dd r' \, \mathrm{e}^{(d-2)A}\delta V_{\Delta_1,\Delta_2}(r') \psi_n^{(1)}(r') \psi_m^{(2)}(r') \right. \notag \\
& \left. \qquad \qquad \qquad \qquad \times \int \! \dd^d y' \, \sqrt{-g_{\mathrm{AdS}_d}} G_{\Delta_{1,n},d} (x_1,y') G_{\Delta_{2,m},d} (x_2,y') \, \, \right] \,.
\label{boundary}
\end{align}
The first term just represents a sum of conformal blocks or conformal partial waves because the propagator $G_{\Delta_{n},d}$ is equal to $f_\partial (\Delta_n, \eta)$ up to normalization. We can also confirm this by Casimir equation.
By using a following relation, 
\begin{equation}
K_{\Delta,d+1} (r,\vec{x},w,\vec{x},w) = \frac{ \Gamma (\Delta ) (2\Delta -d)}{\pi ^{d/2} \Gamma (\Delta -d/2)} \left( \frac{\cosh r}{w}\right)^{\Delta} \,,
\end{equation}
the terms in front of the parenthesis in \eqref{boundary} can be converted into bulk-to-boundary propagators.
Then, the two-point function is decomposed into geodesic Witten diagrams as Fig. \ref{fig:boundary}.

The second term represents an anomalous dimension of operators on the boundary or defect. In the Feynman diagram language, this term corresponds to a mass shift.
Similar terms have appeared in geodesic Witten diagrams without boundary or defect \cite{Hijano} but there is difference.
In the case without boundary or defect, four-point functions are non-trivial and decomposed into geodesic Witten diagrams.
For $\Delta_1+\Delta_2-\Delta_3-\Delta_4 \in 2 \mathbb{Z} $ in s-channel, anomalous dimensions of operators $\mathcal{O}_i$ appear \cite{ano1,ano2}. 
On the other hand, in the presence of the boundary or defect, the anomalous dimension is associated with boundary operators $\hat{\mathcal{O}}$ not ambient operators $\mathcal{O}$.
In conclusion we confirmed that the two-point function can be decomposed into conformal blocks in the boundary channel, and the anomalous conformal dimension terms appeared because the boundary operators receive corrections to their conformal dimensions.

\section{Examples}
\label{sec4}

In the previous section we obtain two different decompositions \eqref{ambient} and \eqref{boundary}.
In this section we explain how our prescription works using a D3/D5 brane system \cite{KR1,KR2} and also comment on the Takayanagi's proposal \cite{Takayanagi} and the Janus solutions \cite{Janus}.
Correlation functions in the D3/D5 model have been discussed holographically in \cite{DFO,DFGK} and also geodesic Witten diagrams have already obtained in \cite{RZ}.
However we would like to stress that our prescription can concretely determine coefficients in front of geodesic Witten diagrams, appeared in two-point functions.

\begin{figure}[t]
 \begin{minipage}{0.5\hsize}
  \begin{center}
   \begin{tikzpicture}
\fill (0,0) circle (3pt);
\fill[green] (-1.5,0) circle (4pt);
\fill[green] (-3.5,0) circle (4pt);
\node[yshift=2pt,above]  at (-1.3,0) {$\mathcal{O}_1(\vec{x}_1,w_1)$};
\node[yshift=2pt,above]  at (-3.5,0) {$\mathcal{O}_2(\vec{x}_2,w_2)$};
\draw[line width=1pt] (-4.5,0) -- (0,0) -- (2.5,0);
\draw[dotted] (0,0) -- (0,-4);

\draw[dashed] (-3.5,0) arc [start angle = 180, end angle = 360, radius = 1];

\draw[snake it,line width=1pt] (-1.5,0) -- (-2.5,-1);
\draw[snake it,line width=1pt] (-3.5,0) -- (-2.5,-1);
\draw[snake it,line width=1pt] (-2.5,-1) -- (0,-3);
\node at (-0.8,-0.8) {$K_{\Delta_1,d+1}$};
\node at (-3.8,-0.8) {$K_{\Delta_2,d+1}$};
\node at (-2.1,-2.3) {$G_{\Delta_N,d+1}$};
\fill (0,-3) circle (3pt);
\node at (0,-4.5) {D5 brane};
\end{tikzpicture}
  \end{center}
  \subcaption{The ambient channel}
 \end{minipage}
 \begin{minipage}{0.5\hsize}
  \begin{center}
   \begin{tikzpicture}
\fill (8,0) circle (3pt);
\fill[green] (6.5,0) circle (4pt);
\fill[green] (4.5,0) circle (4pt);
\node[yshift=2pt,above]  at (6.7,0) {$\mathcal{O}_1(\vec{x}_1,w_1)$};
\node[yshift=2pt,above]  at (4.5,0) {$\mathcal{O}_2(\vec{x}_2,w_2)$};
\draw[line width=1pt] (3.5,0) -- (8,0) -- (10.5,0);
\draw[dotted] (8,0) -- (8,-4);

\draw[dashed] (4.5,0) arc [start angle = 180, end angle = 310, radius = 3.5];
\draw[dashed] (6.5,0) arc [start angle = 180, end angle = 360, radius = 1.5];

\draw[snake it,line width=1pt] (6.5,0) -- (8,-1.5);
\draw[snake it,line width=1pt] (4.5,0) -- (8,-3.5);
\draw[snake it,line width=1pt] (8,-3.5) --  (8,-1.5);
\node at (6.8,-1.1) {$K_{\Delta_1,d+1}$};
\node at (5.9,-2.3) {$K_{\Delta_2,d+1}$};
\node at (8.8,-2.5) {$G_{\Delta_n,d}$};
\node at (8,-4.5) {D5 brane};
\end{tikzpicture}
  \end{center}
  \subcaption{The boundary channel}
  \label{fig:two}
 \end{minipage}
\caption{Two distinct decompositions in the D3/D5 brane system.}
\label{fig:D3D5}
\end{figure}
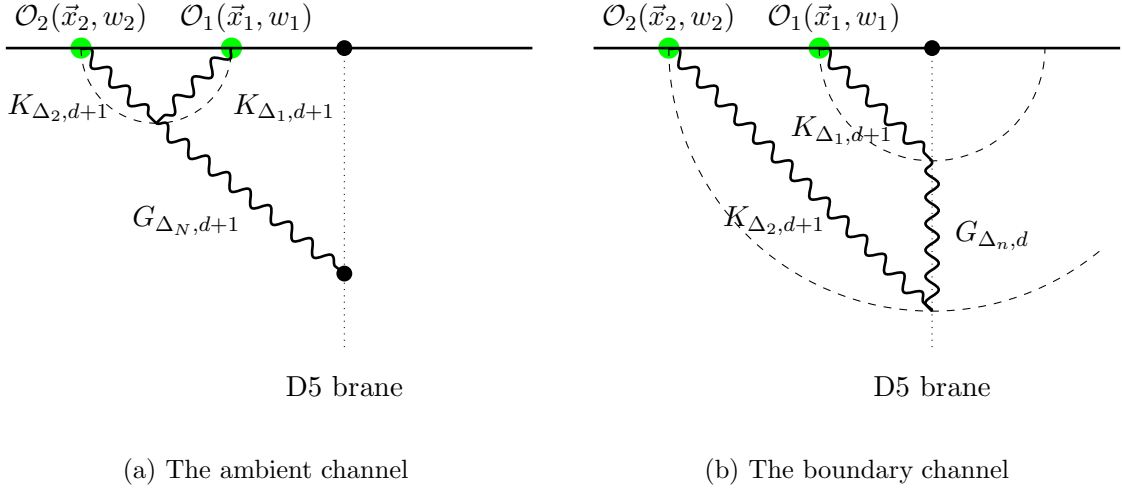

The D3/D5 brane system is constructed from a stuck of $N$ D3-branes and a stuck of  $N_{\mathrm{f}}$ D5-branes.
The D3-branes extend $0,1,2,3$-directions, while the D5-branes extend $0,1,2,4,5,6$-directions.
We assume that $N$ and 't Hooft coupling $\lambda$ are large and the number of flavour branes $N_\mathrm{f}$ is much smaller than $N$, $N_\mathrm{f}/N \ll 1$, in order to treat the D5-branes as probes.
The dual field theory of this model is the $d=4$, $\mathcal{N}=4$ supersymmetric gauge theory coupled to flavour fields on a defect extending $0,1,2$-directions.
The action of the bulk constructed from type IIB supergravity action for the D3-branes $S_{\mathrm{IIB}}$, a DBI action $S_{\mathrm{DBI}}$, and a Chern-Simons action  $S_{\mathrm{CS}}$, for D5-branes,
\begin{equation}
S=S_{\mathrm{IIB}}+S_{\mathrm{DBI}}+S_{\mathrm{CS}} \,.
\end{equation}
The type IIB supergravity action compactified on $S^5$ sphere contains 
\begin{equation}
S_{\mathrm{bulk}}= -N^2 \int \! \dd ^{5}X \, \sqrt{-g_0} \left[ \sum_I \left( \frac{1}{2} (\partial_M \phi_I )^2 +\frac{M^2}{2}\phi_I^2 \right)  +\sum_{I,J,K}\alpha_3 \phi_I \phi_J \phi_K +\cdots \right]
\end{equation}
where $g_0$ is AdS$_5$ metric and $\phi_I$ represents bulk fields.
In this section, we do not care about coefficients of interaction terms which is independent on $N$, $\lambda$ and $N_\mathrm{f}$ and do not give explicit expressions for such coefficients.
The action for the defect is 
\begin{equation}
\begin{split}
S_{\mathrm{defect}}
&= -N \lambda^{1/2}N_\mathrm{f}\int \dd ^{4}x  \, \sqrt{-g_{\mathrm{AdS}_4}}  \\
& \qquad \qquad \times \left[ \sum_I \frac{1}{2}(\partial_i \psi_I)^2  + \sum_{m,n\geq 0} \sum_{I,J}\beta_{m,n} \phi_{I_1}\cdots \phi_{I_m} \psi_{J_1}\cdots \psi_{J_n} +\cdots \right]
\end{split}
\end{equation}
where $\psi_I$ defect scalar fields.
To canonically normalize kinetic terms, we define rescaled fields,
\begin{equation}
\phi' :=N\phi \,, \quad \psi' :=\sqrt{NN_\mathrm{f}\lambda^{1/2}}\psi \,,
\end{equation}
then the actions become schematically 
\begin{align}
S_{\mathrm{bulk}} &= - \int \! \dd ^{5}X \, \sqrt{-g_{5}} \left( \frac{1}{2} (\partial_M \phi')^2 +\frac{M^2}{2}\phi'^2  +\frac{\alpha_3}{N} \phi'^3 +\cdots \right)\,,  \\
S_{\mathrm{defect}} &= -\int \dd ^{4}x  \, \sqrt{-g_{4}} \left( \frac{1}{2}(\partial_m \psi')^2  + \sum_{m,n\geq 0} \beta_{m,n} N^{1-m-n/2}\lambda^{1/2-n/4}N_\mathrm{f}^{1-n/2}  \phi'^m \psi'^n +\cdots \right) \,.
\end{align}
The potential is determined from the interaction term in $S_{\mathrm{defect}}$,
\begin{align}
2 \mathrm{e}^{-2A}\varepsilon \, \delta V_{\Delta_1,\Delta_2} (r)
&= - \frac{\delta^2 S_{\mathrm{defect}}}{\delta \phi_{\Delta_1} \delta \phi_{\Delta_2}} \notag \\
&=\beta_{2,0}\sqrt{\lambda}\frac{N_\mathrm{f}}{N}\delta (r) \,.
\label{deltapot}
\end{align}
This potential is proportional to the small parameter $N_{\mathrm{f}}/N$ and the delta function as expected.
Inserting this potential \eqref{deltapot} into obtained decompositions \eqref{ambient} and \eqref{boundary}, we recover the same geodesic Witten diagrams in \cite{RZ}.
However, our prescription determines also coefficients coming from vertices.
The ambient channel decomposition involves the bulk $\phi^3$ interaction and the defect $\phi \psi$ interaction.
Each vertices have orders $1/N$ and $\sqrt{\lambda} N_\mathrm{f}$ and totally gives an order $\sqrt{\lambda}N_\mathrm{f}/N$ as expected.
On the other side, the boundary channel decomposition involves the defect $\phi \psi$ interaction.
The order of the vertex is $(\sqrt{\lambda}N_\mathrm{f}/N)^{1/2}$ as expected.
Our prescription works well in the D3/D5 brane system. See Fig. \ref{fig:D3D5}.

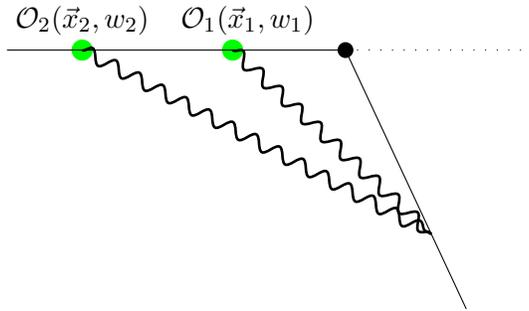
\begin{figure}[t]
\begin{center}
\begin{tikzpicture}
\fill (0,0) circle (3pt);
\fill[green] (-1.5,0) circle (4pt);
\fill[green] (-3.5,0) circle (4pt);
\node[yshift=2pt,above] (O1) at (-1.3,0) {$\mathcal{O}_1(\vec{x}_1,w_1)$};
\node[yshift=2pt,above] (O1) at (-3.5,0) {$\mathcal{O}_2(\vec{x}_2,w_2)$};
\draw (-4.5,0) -- (0,0);
\draw[loosely dotted]  (0,0) -- (2.5,0);
\draw (0,0) -- (295:3.8);

\draw[snake it,line width=1pt] (-1.5,0) -- (295:2.7);
\draw[snake it,line width=1pt] (-3.5,0) -- (295:2.7);
\end{tikzpicture}
\caption{A leading correction of two-point functions in the Takayanagi's proposal.}
\label{fig:takayana}
\end{center}
\end{figure}

Comments on other examples are in order.
The Takayanagi's proposal is a holographic model of boundary CFTs.
In this proposal, an additional boundary is introduced in the bulk and the spacetime is restricted in the region $r_\ast <r< \infty$ (See Fig. \ref{fig:takayana}).
Matter fields are localized on the additional boundary.
When the backreaction coming from the boundary can be ignored, the bulk is described by pure AdS.
The main difference between the Takayanagi's proposal and the D3/D5 model is spacetime is terminated at $r=r_\ast$ or not.
Then if we restrict the integration region in \eqref{2pt} appropriately, we can apply our prescription to the Takayanagi's proposal. 

Finally we comment on the Janus solutions.
In previous our paper \cite{KS}, we treated the $d=4$ Janus solution as an example and an extension to different operators are straightforward. 
Hence we give a brief comment on differences between the D3/D5 model and the Janus solutions here.
The dual CFTs of the Janus solutions are so called interface CFTs and the coupling constant jumps at the interface.
This is achieved by a nontrivial profile of the dilaton field.
The potential of the D3/D5 system is localized on the D5 branes and is proportional to the delta function while that of the Janus is not localized and spreads in whole of AdS.
Since interaction terms also comes from kinetic terms coupled to the dilaton, small modifications are needed.
Our prescription also works nicely in Janus solutions.

\section{Conclusion and Discussion}
\label{con}

In the presence of boundaries or defects in CFTs, two-point functions are still nontrivial and are functions of the conformal cross-ratio.
Furthermore the two-point functions do not vanish even if conformal dimensions are different.
In BCFTs or dCFTs, the two-point functions can be decomposed into conformal blocks or conformal partial waves in two distinct ways; ambient channel decomposition and boundary channel decomposition.
The holographic dual of conformal blocks in dCFT was initiated by Rastelli and Zhou \cite{RZ}.
They discussed a very simple model, a D3/D5 brane model, and obtained geodesic Witten diagrams with general conformal dimensions.
In our previous work \cite{KS}, we gave decompositions of two-point functions into geodesic Witten diagrams in general holographic setups with boundaries or defects.
Our work includes results in \cite{RZ} as a special case but considered only the case with two identical operators.
In this paper we generalized our previous paper to a situation where two operators are different and have different conformal dimensions.

The leading correction of the two-point functions is given by two bulk-to-boundary propagators coupled to a source term which represents an existence of boundaries or defects \eqref{2pt}.
We derived the ambient channel decomposition \eqref{ambient} and the boundary channel decomposition from \eqref{2pt}.
Note that these two different decompositions are equivalent since we derived them from \eqref{2pt}.
In the boundary channel, the decomposition of the two-point functions includes not only
conformal partial waves but also anomalous dimensions if the conformal dimensions satisfy $\Delta_1-\Delta_2 \in \mathbb{Z}$.

To present that our prescription works, we applied it to a D3/D5 brane system.
Defect matter fields localize on the D5 brane and the potential is proportional to a delta function. 
When the probe D5-branes sit on $r=0$, our geodesic Witten diagrams reduce to that of \cite{RZ}.
Recently, two-point functions in dCFT dual to the D3/D5 system with flux have been discussed \cite{deLeeuw:2017dkd}.
It would be nice to compare their work with our prescription.
In addition, we commented on other examples, the Takayanagi's proposal and the Janus soutions.
We confirmed that our prescription works well in both examples.

In this paper we only considered scalar operators to simplify our discussion. 
Geodesic Witten diagrams with an external spin was initiated in \cite{NT} and extended to arbitrary external and internal spins in \cite{Castro,Dyer,Hideki}.
It is possible to generalize our discussion to two-point functions with spins.
Our starting equation \eqref{2pt} involves bulk-to-boundary propagators, $K_{\Delta,d+1}$. If one wants to consider two-point functions with spins, one should replace the bulk-to-boundary propagators without spins by that with spins.
However, we need an equation similar to \eqref{39} for the ambient channel decomposition and a decomposition similar to \eqref{224} for the boundary channel decomposition.
These are not easy tasks because a decomposition of two-point functions may include various representations.
We leave this problem as a future work.

\section*{Acknowledgements}
The author would like to thank Andreas Karch, Tatsuma Nishioka and Zhu Rui Dong for fruitful discussions.
The work is supported by the Grant-in-Aid for Japan Society for the Promotion of Science Fellows, No. 16J01567.

\appendix

\bibliographystyle{JHEP}
\bibliography{draft}

\providecommand{\href}[2]{#2}\begingroup\raggedright\begin{thebibliography}{10}

\bibitem{KS}
A.~Karch and Y.~Sato, {\it {Boundary Holographic Witten Diagrams}},  {\em JHEP}
  {\bf 09} (2017) 121, \href{http://xxx.lanl.gov/abs/1708.01328}{{\tt
  1708.01328}}.

\bibitem{Witten}
E.~Witten, {\it {Anti-de Sitter space and holography}},  {\em Adv. Theor. Math.
  Phys.} {\bf 2} (1998) 253--291,
  \href{http://xxx.lanl.gov/abs/hep-th/9802150}{{\tt hep-th/9802150}}.

\bibitem{Hijano}
E.~Hijano, P.~Kraus, E.~Perlmutter, and R.~Snively, {\it {Witten Diagrams
  Revisited: The AdS Geometry of Conformal Blocks}},  {\em JHEP} {\bf 01}
  (2016) 146, \href{http://xxx.lanl.gov/abs/1508.00501}{{\tt 1508.00501}}.

\bibitem{McAvity}
D.~M. McAvity and H.~Osborn, {\it {Conformal field theories near a boundary in
  general dimensions}},  {\em Nucl. Phys.} {\bf B455} (1995) 522--576,
  \href{http://xxx.lanl.gov/abs/cond-mat/9505127}{{\tt cond-mat/9505127}}.

\bibitem{ADFK}
O.~Aharony, O.~DeWolfe, D.~Z. Freedman, and A.~Karch, {\it {Defect conformal
  field theory and locally localized gravity}},  {\em JHEP} {\bf 07} (2003)
  030, \href{http://xxx.lanl.gov/abs/hep-th/0303249}{{\tt hep-th/0303249}}.

\bibitem{KR1}
A.~Karch and L.~Randall, {\it {Open and closed string interpretation of SUSY
  CFT's on branes with boundaries}},  {\em JHEP} {\bf 06} (2001) 063,
  \href{http://xxx.lanl.gov/abs/hep-th/0105132}{{\tt hep-th/0105132}}.

\bibitem{KR2}
A.~Karch and L.~Randall, {\it {Locally localized gravity}},  {\em JHEP} {\bf
  05} (2001) 008, \href{http://xxx.lanl.gov/abs/hep-th/0011156}{{\tt
  hep-th/0011156}}.

\bibitem{Takayanagi}
T.~Takayanagi, {\it {Holographic Dual of BCFT}},  {\em Phys. Rev. Lett.} {\bf
  107} (2011) 101602, \href{http://xxx.lanl.gov/abs/1105.5165}{{\tt
  1105.5165}}.

\bibitem{Janus}
D.~Bak, M.~Gutperle, and S.~Hirano, {\it {A Dilatonic deformation of AdS(5) and
  its field theory dual}},  {\em JHEP} {\bf 05} (2003) 072,
  \href{http://xxx.lanl.gov/abs/hep-th/0304129}{{\tt hep-th/0304129}}.

\bibitem{RZ}
L.~Rastelli and X.~Zhou, {\it {The Mellin Formalism for Boundary CFT$_d$}},
  {\em JHEP} {\bf 10} (2017) 146,
  \href{http://xxx.lanl.gov/abs/1705.05362}{{\tt 1705.05362}}.

\bibitem{Liendo}
P.~Liendo, L.~Rastelli, and B.~C. van Rees, {\it {The Bootstrap Program for
  Boundary CFT$_d$}},  {\em JHEP} {\bf 07} (2013) 113,
  \href{http://xxx.lanl.gov/abs/1210.4258}{{\tt 1210.4258}}.

\bibitem{DFGK}
O.~DeWolfe, D.~Z. Freedman, S.~S. Gubser, and A.~Karch, {\it {Modeling the
  fifth-dimension with scalars and gravity}},  {\em Phys. Rev.} {\bf D62}
  (2000) 046008, \href{http://xxx.lanl.gov/abs/hep-th/9909134}{{\tt
  hep-th/9909134}}.

\bibitem{Hogervorst}
M.~Hogervorst, {\it {Crossing Kernels for Boundary and Crosscap CFTs}},
  \href{http://xxx.lanl.gov/abs/1703.08159}{{\tt 1703.08159}}.

\bibitem{HH}
C.~P. Herzog and K.-W. Huang, {\it {Boundary Conformal Field Theory and a
  Boundary Central Charge}},  {\em JHEP} {\bf 10} (2017) 189,
  \href{http://xxx.lanl.gov/abs/1707.06224}{{\tt 1707.06224}}.

\bibitem{ano1}
H.~Liu, {\it {Scattering in anti-de Sitter space and operator product
  expansion}},  {\em Phys. Rev.} {\bf D60} (1999) 106005,
  \href{http://xxx.lanl.gov/abs/hep-th/9811152}{{\tt hep-th/9811152}}.

\bibitem{ano2}
E.~D'Hoker, S.~D. Mathur, A.~Matusis, and L.~Rastelli, {\it {The Operator
  product expansion of N=4 SYM and the 4 point functions of supergravity}},
  {\em Nucl. Phys.} {\bf B589} (2000) 38--74,
  \href{http://xxx.lanl.gov/abs/hep-th/9911222}{{\tt hep-th/9911222}}.

\bibitem{DFO}
O.~DeWolfe, D.~Z. Freedman, and H.~Ooguri, {\it {Holography and defect
  conformal field theories}},  {\em Phys. Rev.} {\bf D66} (2002) 025009,
  \href{http://xxx.lanl.gov/abs/hep-th/0111135}{{\tt hep-th/0111135}}.

\bibitem{deLeeuw:2017dkd}
M.~de~Leeuw, A.~C. Ipsen, C.~Kristjansen, K.~E. Vardinghus, and M.~Wilhelm,
  {\it {Two-point functions in AdS/dCFT and the boundary conformal bootstrap
  equations}},  {\em JHEP} {\bf 08} (2017) 020,
  \href{http://xxx.lanl.gov/abs/1705.03898}{{\tt 1705.03898}}.

\bibitem{NT}
M.~Nishida and K.~Tamaoka, {\it {Geodesic Witten diagrams with an external
  spinning field}},  {\em PTEP} {\bf 2017} (2017), no.~5 053B06,
  \href{http://xxx.lanl.gov/abs/1609.04563}{{\tt 1609.04563}}.

\bibitem{Castro}
A.~Castro, E.~Llabres, and F.~Rejon-Barrera, {\it {Geodesic Diagrams,
  Gravitational Interactions \& OPE Structures}},  {\em JHEP} {\bf 06} (2017)
  099, \href{http://xxx.lanl.gov/abs/1702.06128}{{\tt 1702.06128}}.

\bibitem{Dyer}
E.~Dyer, D.~Z. Freedman, and J.~Sully, {\it {Spinning Geodesic Witten
  Diagrams}},  \href{http://xxx.lanl.gov/abs/1702.06139}{{\tt 1702.06139}}.

\bibitem{Hideki}
H.-Y. Chen, E.-J. Kuo, and H.~Kyono, {\it {Anatomy of Geodesic Witten
  Diagrams}},  {\em JHEP} {\bf 05} (2017) 070,
  \href{http://xxx.lanl.gov/abs/1702.08818}{{\tt 1702.08818}}.

\end{thebibliography}\endgroup

\end{document}